# Unusual spin properties of InP wurtzite nanowires revealed by Zeeman splitting spectroscopy

D. Tedeschi,[‡] M. De Luca,[‡,+] P. E. Faria Junior,[#] A. Granados del Águila,[†,⊥] Q. Gao,[^] H. H. Tan,[^] B. Scharf,[#,×] P. C. M. Christianen,[†] C. Jagadish,[^] J. Fabian[#] and A. Polimeni[‡,*]

[‡]Dipartimento di Fisica, Sapienza Università di Roma, 00185 Roma, Italy

[†]High Field Magnet Laboratory (HFML – EMFL), Radboud University, NL-6525 ED Nijmegen, The Netherlands

[#]Institute for Theoretical Physics, University of Regensburg, 93040 Regensburg, Germany

[^]Department of Electronic Materials Engineering, Research School of Physics and Engineering, The Australian National University, Canberra, ACT 2601, Australia

In this study, we present a complete experimental and theoretical investigation of the fundamental exciton Zeeman splitting in wurtzite InP NWs. We determined the exciton gyromagnetic factor, $g_{exc}$, by magneto-photoluminescence spectroscopy using magnetic fields up to 29 T. We found that $g_{exc}$ is strongly anisotropic with values differing in excess of 50% between the magnetic field oriented parallel and perpendicular to the nanowire long axis. Furthermore, for magnetic fields oriented along the nanowire axis, $g_{exc}$ is nearly three times larger than in bulk zincblende InP and it shows a marked sublinear dependence on the magnetic field, a common feature to other non-nitride III-V wurtzite nanowires but not properly understood. Remarkably, this nonlinearity originates from only one Zeeman branch characterized by a specific type of light polarization. All the experimental findings are modelled theoretically by a robust approach combining the *k·p* method with the envelope function approximation and including the electron-hole interaction. We revealed that the nonlinear features arise due to the coupling between Landau levels pertaining to the A (heavy-hole like) and B (light-hole like) valence bands of the WZ crystal structure. This general behavior is particularly relevant for the understanding of the spin properties of several wurtzite nanowires that host the set for the observation of topological phases potentially at the base of quantum computing platforms.

The value of the gyromagnetic factor, *g*-factor, of electrons and holes in solids dictates their spin response to applied magnetic fields ($\vec{B}$) and reflects the strength of relativistic effects acting on carriers[1]. In recent years, the investigation of the spin properties (such as value and anisotropy of the *g*-factor) of semiconducting nanowires (NWs) has attracted a lot of interest because these properties determine the conditions to detect Majorana fermions that are the main candidates for topological quantum-computation[2]. While the spin properties of III-V NWs with a zincblende (ZB) crystal structure are well known due to the similarity with the same material in the bulk form, this is not the case of NWs with the wurtzite (WZ) crystal phase, as WZ does not exist in the bulk counterpart of many important non-nitrides III-V semiconductors, such as GaAs, InP and InAs.

In all materials, the investigation of *g*-factors has also a fundamental importance because the comparison between the values experimentally determined and those theoretically calculated offers a stringent test for band structure calculations. In particular, the dependence of the *g*-factor on the relative direction between $\vec{B}$ and specific crystallographic directions of the lattice can provide insightful information about the symmetry characteristics of the underlying electronic structure[3,4,5]. This fact manifests in expressions of the Zeeman splitting (ZS) that may differ considerably depending on the experimental configuration employed. In the case of $C_{6V}$ symmetry pertinent to the WZ NWs investigated in this work, the energy of the Zeeman split components of the exciton are given by[6,7,8]

$$\Gamma_5^{\pm} - DS_F = S \cdot \mu_B |g_{e,\parallel} - g_{h,\parallel}| B, \quad \text{for } \vec{B} \parallel \vec{q} \parallel \hat{c}; \tag{1}$$

$$\Gamma_{5/6}^{\pm} - DS_V = S \cdot \sqrt{\Delta_{56}^2 + \mu_B^2 g_{e,\perp}^2 B^2}, \quad \text{for } \vec{B} \perp \vec{q} \parallel \hat{c}, \tag{2}$$

where $S$ is the spin vector component ($S = \pm\frac{1}{2}$), $\mu_B$ is the Bohr magneton, $\vec{q}$ is the emitted photon wavevector, $\hat{c}$ is the [0001] axis of the WZ lattice (corresponding to the axial direction in NWs, as depicted in the insets in Figure 1) and $\Delta_{56}$ is the electron-hole exchange energy separating spin-allowed $\Gamma_5$ (bright) and spin-forbidden $\Gamma_6$ (dark) excitons within the dipole approximation[9]. $\Gamma_{5/6}$ is the mixed state resulting when an external perturbation (*e.g*, a magnetic field) breaks the WZ crystal symmetry[6,7,8]. Finally, $DS_{F/V}$ is the exciton diamagnetic shift relative to the Faraday (F, $\vec{B} \parallel \vec{q}$) and Voigt (V, $\vec{B} \perp \vec{q}$) configurations. "e" and "h" subscripts refer to electron and hole related parameters, respectively. Therefore, a linear and quasi-linear dependence on $B$ is expected for $\vec{B}$ parallel and perpendicular to $\vec{q}$, respectively, each being ruled by a different *g*-factor (labeled in Eq. (1) as $g_\parallel$ and $g_\perp$, respectively). Departures from these behaviors may occur whenever *g* is field-dependent, a case that was most observed in quantum wells and ascribed to the coupling between heavy- and light-hole subbands[10,11,12,13,14]. No similar effects were reported in bulk materials, while a *B*-dependent *g* was first observed in NWs featuring the WZ crystal phase (InP, GaAs, and InGaAs)[8,15,16] but never modeled. Since the size of those WZ NWs rules out carrier quantum confinement effect, the observation of the magnetic field dependence of the *g*-factor has to be attributed exclusively to the WZ phase, which is not found in bulk, and not to the NW shape/size. Finally, the observation of Majorana-like behavior was reported in InAs WZ NWs[2,17], where a field-dependent *g*-factor of carriers could be potentially relevant to unveil the associated phenomenon.

In this work, we measured by magneto-photoluminescence (PL) and modelled with the $k\cdot p$ method the exciton Zeeman splitting in WZ NWs. For this kind of combined approach, we focus on InP NWs, which is the NW III-V system where most of the main band structure properties have been best established to date.[18] Magnetic fields as high as 29 T were used in two different experimental geometries: $\vec{B} \parallel \vec{q} \parallel \hat{c}$ (Faraday) and $\vec{B} \perp \vec{q} \parallel \hat{c}$ (Voigt). In the latter case, the exciton ZS data follow Eq. (2) with $g_{e,\perp}$ in excellent agreement with our theoretical expectations based on a multi-band $k \cdot p$ theory involving a novel multiband $k \cdot p$ Hamiltonian. In the Faraday configuration, instead, the exciton ZS exhibits a strongly nonlinear field-dependence and only at low fields ($B$<5 T) the calculated $g$-factors *via* the $k \cdot p$ model can account for the experimental results. We explain the experimental data at large magnetic fields by developing a general description that takes into account the spatial dependence of the vector potential in the $k \cdot p$ Hamiltonian. In particular, we find that the admixing between wavefunctions of the A (heavy-hole, HH, like) and B (light-hole, LH, like) valence bands (VBs) characteristic of the WZ lattice leads to the observed nonlinear behavior of the exciton ZS for $\vec{B} \parallel \vec{q} \parallel \hat{c}$. It is noteworthy that this non linearity is borne by just one ($\Gamma_5^+$) of the two Zeeman split components. We show that this finding can be traced back to the interplay between the symmetry of the VB wavefunctions and their spin. Finally, we obtain a quantitative agreement with the experimental data for all field geometries by including the exciton effects.

The InP NW samples investigated in this work were grown by selective-area-epitaxy (SAE) and by the vapor-liquid-solid (VLS) method. In both cases, the NWs exhibit nearly pure WZ phase and perfect vertical alignment to their respective substrates. Details about the growth and the optical properties of SAE and VLS NWs can be found in Refs 8 and 19, respectively. The insets in Figure 1 show the scanning electron microscopy images of the SAE NWs, on which the measurements presented in this paper were performed. The NWs have an average diameter and height of 650 nm and 5 μm, respectively, thus allowing us to assume these NWs as bulk-like materials where quantum confinement is negligible. We also stress that the experimental data we reported here are independent of the NW morphological characteristics, such as diameter, length, as well as of the growth technique employed (SAE or VLS). For PL measurements under magnetic field, the samples were placed in a water-cooled Bitter magnet at $T$=4.2 K using a bath cryostat. PL was excited by a frequency-doubled Nd:YVO$_4$ laser ($\lambda$=532 nm) focused using a long focal length objective (spot diameter ~10 μm), collected by the same objective, dispersed by a 0.30 m monochromator, and detected by a liquid nitrogen-cooled Si CCD. Polarization of the emitted light was analyzed using a combination of a liquid crystal variable retarder working as a quarter-wave plate and a linear polarizer in order to make all measurements insensitive to the polarization response of the optical set-up. Theoretical investigation was performed under the framework of the *k·p* method, using the recently reported Hamiltonian and InP WZ parameters of Ref. 20. For the linear Zeeman splitting regime, we calculated the *g*-factors using a perturbative technique similar to the conventional Roth, Lax, and Zwerdling approach[21]. To capture the nonlinear features, we investigate the single-particle Landau level spectra including the spatial dependence of the vector potential within the envelope function approximation[22]. The resulting Hamiltonian is then solved numerically using the finite difference method[23]. Finally, for a more realistic comparison to the experimental exciton Zeeman splitting we included the electron-hole Coulomb effects *via* the Bethe-Salpeter equation[24].

The low-temperature (4.2 K) magneto-PL spectra of the WZ InP NWs, showing the band-gap free-exciton (FE) energy region, are displayed in Figure 1 from 0 to 29 T. Panel (a) refers to measurements in which $\vec{B}$ was directed parallel to the NW $\hat{c}$ axis and the luminescence was collected along the $\hat{c}$ direction ($\vec{B} \parallel \hat{c} \parallel \vec{q}$), see inset. Γ$_5$ denotes the symmetry of the bright exciton formed by the conduction Γ$_{7C}$ and valence Γ$_{9V}$ band states of the WZ lattice[25,26]. The spectra are shown only for PL detected selecting σ$^+$ circularly polarized light for clarity reasons (measurements with σ$^-$ detection were also taken, and they are shown in the Supplemental

Material, SM, section I). At zero field, the FE lineshape exhibits a doublet structure caused by the scattering of the excitons with donor impurities[27], as discussed in Ref. 8. These two components [named $\Gamma_5$(U) and $\Gamma_5$(L) at upper and lower energy, respectively] are analyzed separately and will be regarded as two distinct exciton distributions. Eventually the relevant physical quantities are evaluated as an average over these two distributions, where appropriate. With increasing $B$, $\Gamma_5$(U) and $\Gamma_5$(L) both split into $\Gamma_5^\pm$ Zeeman split states and shift in energy. The corresponding diamagnetic shift (extensively discussed in Ref. 8) has been subtracted to each $B \neq 0$ spectrum. This allows us to highlight the Zeeman splitting of the U and L components. We point out that in this configuration the emitted photons are circularly dichroic and the Zeeman-split levels can be discerned by suitable polarization optics. The $\Gamma_5^\pm$ components are highlighted by solid lines in Figure 1 (a).

Figure 1 (b) shows the $T$=4.2 K magneto-PL spectra for $\vec{B}$ directed perpendicular to the NW $\hat{c}$ axis, see inset, while the luminescence is collected along the $\hat{c}$ direction ($\vec{B} \perp \hat{c} // \vec{q}$). Under this geometry, $\Gamma_5$ and $\Gamma_6$ excitons are mixed by the field and the resultant state is usually indicated as $\Gamma_{5/6}$[3,25,26]. As in panel (a), each $B \neq 0$ spectrum has been redshifted by the corresponding diamagnetic shift[8]. Due to the relative orientation between the field and the luminescence $\vec{q}$ vector, no PL circular dichroism is observed[3,7,9,25] and no circular polarization filtering can be exploited in this case to make the ZS more visible. However, the good spectral resolution along with the small full width at half maximum of the PL bands makes it still clearly visible. Indeed, the PL peaks corresponding to the Zeeman-split $\Gamma_{5/6}^\pm$ branches are highlighted by solid lines in Figure 1 (b). A bound exciton (BE) is also observed for $B$>10 T[8].

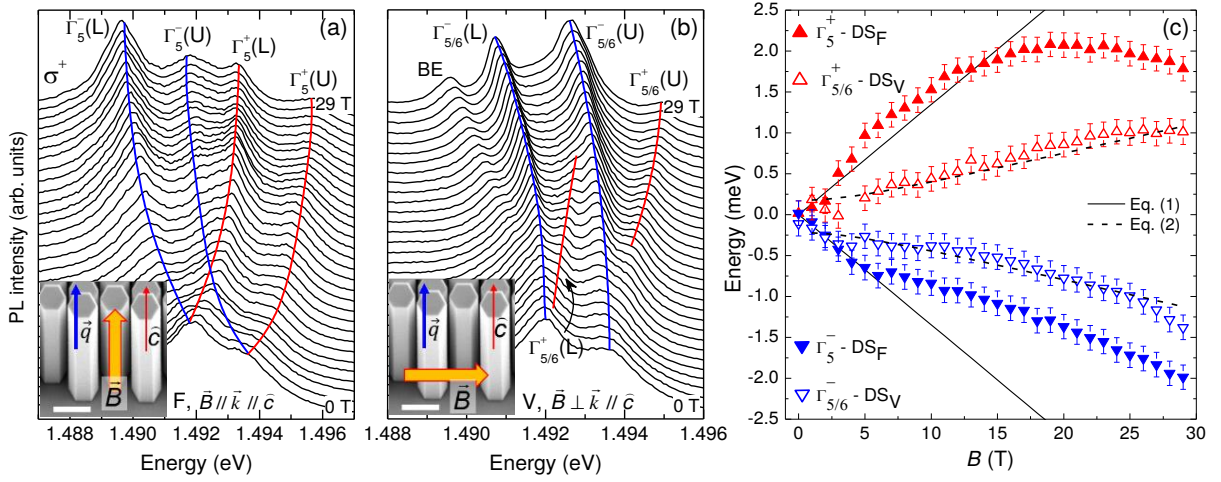

**Figure 1**. (a) Low-temperature ($T$=4.2 K) photoluminescence spectra of WZ InP NWs from 0 to 29 T (with increasing field from 0 to 29 T in step of 1 T from bottom to top). The relative orientations between magnetic field, emitted photon wavevector $\vec{q}$, and WZ $\hat{c}$ axis in the Faraday geometry are $\vec{B}/\!/\vec{q}/\!/\hat{c}$ (inset). The spectra were filtered by $\sigma^+$ circular polarization. The different components are indicated and highlighted by colored lines. The energy axis of each $B\neq 0$ T spectrum has been corrected by the corresponding diamagnetic shift. (b) Same as (a) for the Voigt $\vec{B}\perp\vec{q}/\!/\hat{c}$ geometry.

Insets: scanning electron microscopy images of the SAE NWs (scale bar is 1 μm) on which the relevant vector directions are superimposed to illustrate the experimental geometry.

(c) Energy of the various free-exciton components displayed in panels (a) and (b) after subtraction of the pertinent diamagnetic shift ($DS_i$; i = F,V). The corresponding geometrical configurations are indicated by the subscripts F and V, which stand respectively for the $\vec{B}/\!/\vec{q}/\!/\hat{c}$ (Faraday, full triangles) and $\vec{B}\perp\vec{q}/\!/\hat{c}$ (Voigt, empty triangles) configurations. Solid lines are the theoretical values obtained via Equation (1) with $|g_{e,\parallel} - g_{h,\parallel}| = 4.66$ for $\vec{B}/\!/\vec{q}/\!/\hat{c}$. Dashed lines are fittings to the data via Equation (2) with only $\Delta_{56}$ (=0.3±0.2 meV) as a free parameter, being $g_{e,\perp} = 1.29$ the theoretical prediction for $\vec{B}\perp\vec{q}/\!/\hat{c}$.

Figure 1 (c) shows the field evolution of the Zeeman-split states resulting from the spectra displayed in panels (a) and (b). In order to determine the energy of the exciton states for the largest range of fields we resorted to the spectrum second derivative[8]. The data are obtained by subtracting the pertinent diamagnetic shift ($DS_{F/V}$) from the opposite spin components of the exciton. We recall that $DS_{F/V}$ was evaluated previously *via* a perturbative approach[8].

We discuss first the results for the Voigt configuration (open triangles). In this case, the Zeeman-split components can be resolved at all fields only for the $\Gamma_{5/6}(L)$ exciton distribution.

The two $\Gamma_{5/6}^{\pm}$ components shift at the same rate following Equation (2) at all fields. The dashed lines are fittings to the data with only one free parameter: the electron-hole exchange energy $\Delta_{56}$. We set $g_{e,\perp}$=1.29, namely the value calculated *via* the 8×8 $k \cdot p$ bulk Hamiltonian for InP WZ presented in Ref. 20. The details regarding the perturbative calculation of the gyromagnetic factors can be found in section II of SM. It is worth noticing that the experimental value of $\Delta_{56}$=(0.3±0.2) meV is about one order of magnitude larger than in ZB InP[28] indicating a sizably augmented exchange interaction in the hexagonal with respect to the cubic lattice.

In the Faraday configuration, we show the average values of the upper $\Gamma_5(U)$ and lower $\Gamma_5(L)$ exciton distributions (full triangles). In contrast to the Voigt case, the predicted field dependence of the Zeeman splitting, see Equation (1), is only observed for small fields ($B < 5$ T). The solid lines are the theoretical ZS evaluated by Equation (1) with $|g_{e,//} - g_{h,//}|$=4.66 being the value calculated *via* the 8×8 $k \cdot p$ bulk Hamiltonian as in the Voigt configuration, see section II of SM. Remarkably, the two opposite spin components determined experimentally show very different field dependences: $\Gamma_5^-$ is approximately linear, whilst $\Gamma_5^+$ is sublinear pointing to a non-linear ZS, namely a field-dependent *g*-factor. The presence of a field-dependent gyromagnetic factor solely in the Faraday geometry suggests that the valence band states ought to be responsible for such a non-linear ZS. Indeed, only in this configuration the hole gyromagnetic factor plays a role in the ZS, as apparent from Eqs (1) and (2).

In order to describe the magnetic field effects that give rise to the above described nonlinear features in the Faraday geometry, we must move beyond the linear *g*-factor picture described by the simple Eq (1). Thus, we use a more general approach combining the $k \cdot p$ method with the envelope function approximation[22,29]. The total *single-particle* Hamiltonian of the system can be then written as

$$H_{\text{bulk}}\left[\vec{k} \to -i\vec{\nabla} + \frac{e}{\hbar}\vec{A}(\vec{r})\right] + \frac{\mu_B}{2}g_0\vec{\sigma}\cdot\vec{B}, \tag{3}$$

in which $H_{\text{bulk}}$ is the 8×8 *k·p* bulk Hamiltonian for InP WZ given in Ref. 20 with *k*-dependent spin-orbit coupling, $\vec{A}$ is the vector potential, $g_0$ is the bare electron *g*-factor, and $\vec{\sigma}$ are the Pauli matrices written in the bulk basis set, *i. e.*, $\vec{\sigma}\cdot\vec{B}$ is the Zeeman term that takes into account only the spins of the bulk Bloch functions. As discussed, since the NWs have large diameter (>600 nm), we neglect the effects of lateral quantum confinement and treat the NWs as a bulk system[30,31]. For a magnetic field the vector potential has a spatial dependence, which we can choose to be only in one dimension for a suitable gauge (for the Voigt configuration $\vec{B} = B\hat{x} \Rightarrow \vec{A} = B\,y\,\hat{z}$ and for the Faraday configuration $\vec{B} = B\hat{z} \Rightarrow \vec{A} = B\,x\,\hat{y}$, where $\hat{x}, \hat{y}, \hat{z}$ are the units vectors of each spatial direction). As a consequence of this choice in the vector potential form, we have two quantum numbers in $\vec{k}$ space: $k_B$, which is parallel to the magnetic field, and $k_A$, which is parallel to the vector potential. Moreover, the carrier wavevector is two-dimensional ($k_y$-$k_z$ plane for Faraday and $k_x$-$k_z$ plane for Voigt configurations). The choice of a two-dimensional wavevectors is similar to what happens in the description of a quantum well profile,[23,32,33] except that the quantum confinement in our case arises from the spatial dependence of the vector potential. This latter eventually provides the Landau level spectrum of the system. Furthermore, because of the multiband character of the bulk Hamiltonian, we cannot find analytical solutions. However, we can obtain the numerical solutions by using a finite-difference method in real space, as described in Ref. 23. Further details of the calculations can be found in section III of SM. The resulting solution of the Hamiltonian in Eq. (3) gives the Landau level (LL) energies $E_L(k_B, k_A)$ with wavefunctions $\psi_{L,k_B,k_A}$:

$$\psi_{L,k_B,k_A}(\vec{r}) = \frac{e^{i(k_B r_B + k_A r_A)}}{\sqrt{\Lambda}}\sum_{l=1}^{8} f_{L,k_B,k_A,l}(\rho)\, u_l(\vec{r}) \tag{4}$$

where *f*s are the envelope functions, *u*s are the Bloch- functions from the bulk, $L$ labels the Landau level, the coordinate $\rho$ represents the spatial dependence of the vector potential along the specific direction, the summation in $l$ runs over the basis states of the 8 × 8 bulk $k \cdot p$ Hamiltonian, and $\Lambda$ is the area of the system perpendicular to the growth direction. The specific definition of $\rho$, $r_A$, and $r_B$ for the vector potentials in Voigt and Faraday are given in Table II of the Supplemental Material.

In Figure 2 (a) we show the resulting magnetic field dependence of the topmost and bottommost states of the valence and conduction bands, respectively. This dependence is ruled by the linear shift of the pertinent Landau levels LLs (no exciton effects are considered, yet) and by the Zeeman splitting. Therefore, each Landau level contains two branches related to opposite spin configurations as highlighted in Figure 2 (a) by short, colored arrows. These latter specify the carrier spin orientation relative to $\vec{B}$. We stress here that the spin projection is given with respect to the magnetic field direction, regardless of Faraday or Voigt configuration: parallel (antiparallel) spin means spin pointing in the same (opposite) direction of *B*. Solid, colored lines indicate the first Landau level of the topmost valence and bottommost conduction bands. We will show in the following that the first LL of the valence band involves a mixing of HH and LH bands. The dashed, thin (grey) lines indicate higher order Landau levels that do not play a significant role to the excitonic effects.

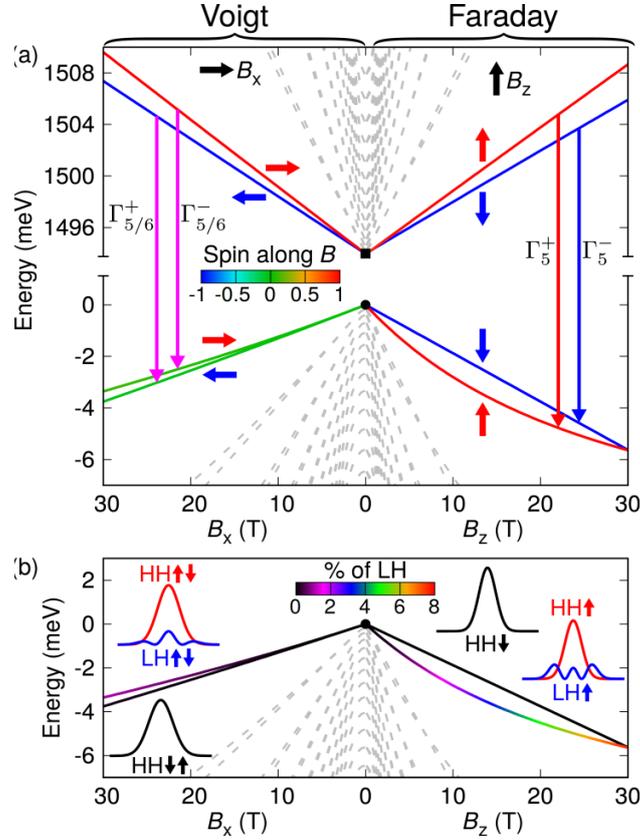

**Figure 2**. (a) Calculated energy dispersion as a function of the magnetic field of the bottom and top most Landau levels for conduction and valence bands, respectively. Solid, colored lines indicate the first Landau level. The color code (and short arrows) indicates the spin projection along the field direction. The vertical arrowed lines connect the levels associated to the measured excitonic transitions labeled according to Figure 1. The dashed, thin (grey) lines indicate the higher order Landau levels that do not play a significant role to the excitonic effects. (b) Calculated energy dispersion as a function of the magnetic field of the top most Landau levels for valence bands as in (a). Here, the color code indicates the percentage of the light hole (LH or B band) character of the pertinent valence band Landau levels. The branches with nonlinear dispersion show a nonzero contribution of LH states that increases with magnetic field. The insets show the probability densities of the corresponding levels at 30 T to highlight the peculiar coupling between HH (0 nodes) and LH (2 nodes) LLs. In Faraday (Voigt) the LH probability density is multiplied by a factor of 5 (15). Along the *x*-axis, values indicate the applied magnetic field in Faraday (right) and Voigt (left) configurations.

*Conduction band*. The situation is relatively simple: in both magnetic field configurations the ZS is linear and the upper (lower) branch has a parallel (antiparallel) spin orientation, which is consistent with positive values of the effective gyromagnetic factors. In general, *g*-factors are positive if the orbital (and spin-orbital) effects are smaller than the bare carrier *g*-factor *g* so that $g_{tot} = g + g_{orbital} > 0$, in which $g_{tot}$ is the total g-factor and $g_{orbital}$ is the orbital (and spin-orbital) contribution. For a more detailed treatment see Eqs (4) and (5) in section II of SM.

*Valence band.* The scenario becomes more complicated, as quite diverse behaviors are observed depending on the field configuration:

- Faraday configuration. The state with antiparallel (parallel) orientation has higher (lower) energy, consistent with a negative gyromagnetic factor. Most importantly, the antiparallel state shows a linear dispersion with respect to the magnetic field, while the parallel state shows a nonlinear dispersion. This is in qualitative agreement with the experimental results presented in Figure 1 (c).

- Voigt configuration. The upper (lower) branch state has a parallel (antiparallel) spin orientation with respect to the magnetic field. We clarify that those branches are green in Figure 2 (a) because the calculated components of the spin along $B$ are very small (0.1 for the upper branch and -0.04 for the lower branch at 30 T), but since the spin projections although small are non zero, they have a direction with respect to $B$, and this direction is represented by the red and blue arrows. We also point out that a small nonzero and nonlinear ZS (~0.4 meV at 30 T) is found by model calculations despite $g_{h,\perp}$ for HH states should be zero[3], showing that the LL approach describes a richer physics beyond the simple linear gyromagnetic factor limit.

Before proceeding to the excitonic ZS, it is important to discuss the origin of the nonlinear dispersion that appears in the valence band in both magnetic field configurations. According to Eq. (4), the total solution of the coupled set of equations contained in Eq. (3) (that includes the spatial dependence of the vector potential) consists of a linear combination of wavefunctions from the bulk energy bands. Such superposition will be different for the different LLs, as different LLs will have different non-zero envelope functions among all those that are present in the general Eq. (4). Specifically regarding the first LL of the valence band, we found that many *f*s of the superposition are nearly zero and mainly HH and LH contributions remain, so that LL consists in a mixing between LH and HH bands. The probability densities of the actual wavefunctions that remain in the superposition of the first LL of the valence band are given in the inset of Figure 2 (b) for $B$=30 T. Let us indicate with $n$ the LLs of the single bands (namely, the HH and LH valence bands). Since $n$ is a quantum number associated with the number of nodes of the wavefunction of the non-interacting HH and LH Landau levels, our calculations show that $n$ is equal to 0 for HH (0 nodes) and 2 for LH (2 nodes); see the inset. Therefore, in Figure 2, the first LL pertinent to the valence band consists of a superposition of different $n$s with different band characters (HH and LH) and thus $n$ is no longer a good quantum number to label the LLs [hence our choice of label L in Eq. (4)].

The fact that the total wavefunction of the first LL of the valence band is a superposition of HH and LH wavefunctions is responsible for the non linearity presented in Figure 2 and in the experimental data. In order to clarify this, we highlight in the main part of Figure 2(b) the calculated contribution of the LH character (pertinent to the band B of WZ) to the different LL branches. Our calculations reveal that only for nonzero contribution of LH the branches are nonlinear [lower (upper) branch for Faraday (Voigt) configuration]. Moreover, such LH contribution increases with magnetic field. At 30 T, for instance, there is ∼ 8% (∼ 1%) of LH character for the Faraday (Voigt) configuration that indicates the mixing between the HH and LH Landau levels.

In summary, the interplay between the WZ symmetry, embedded in the $k \cdot p$ Hamiltonian, and the break of time-reversal symmetry due to the magnetic field, gives rise to the observed LL dispersion presented in Figure 2, which is very peculiar for a bulk-like system. Particularly for the Faraday configuration, the sizeable nonlinear features arise from the LL branch with spin aligned to the magnetic field. Such a spin-dependent nonlinearity can be understood by looking at the leading terms of the Hamiltonian (second order $k \cdot p$ term) reported in the Appendix B of Ref. 20. In a simple picture, the coupling term between spins up connects HH states (pertinent to the band A of WZ) with Landau level $n=0$, to LH states (pertinent to the band B of WZ), with $n=2$ [see Figure 2 (b)]. For the coupling term between spins down, $n=0$ HH states are connected to $n=-2$ LH states. Since $n<0$ are not allowed indices to the LLs, spin down interactions are suppressed. For the Voigt configuration, there is no spin-dependent coupling (both spin components contribute equally) and as a consequence the coupling to the LH state is reduced, hence providing a smaller nonlinear dispersion than the Faraday configuration. This explains why Eq. (2) reproduces quite well the Voigt experimental data while Eq. (1) does not reproduce well the Faraday experimental data.

To illustrate the coupling between HH and LH states in the presence of magnetic field, we can depict the analogy with optical spin orientation in WZ materials (see also section III A of SM)[34]. For light polarization with a particular helicity (for instance, $\sigma^+$) only spin-up transitions from HH to CB are allowed in the electron-electron representation.[35] In order to activate the spin-down transitions, the helicity of the incoming light should be reversed (for instance, to $\sigma^-$). For the LL coupling, an analogous mechanism holds. Considering an applied magnetic field oriented along $z$, only HH and LH LL-states with spin up can be coupled thus giving rise to the branch with nonlinear dispersion. Reversing the orientation of the magnetic field to $-z$ would then allow the coupling between HH and LH LL-states with spin down, obeying the time-reversal symmetry.

We move now to the total excitonic Zeeman splitting observed experimentally obtained in Fig. 1 (c) as energy difference between spin components with opposite helicity. In order to provide a reliable comparison to the experimental results, we included the effects of the exciton interaction following the effective Bethe-Salpeter equation (BSE)[24,36] given by

$$\left[\Delta_{cv}(\vec{k}) - \Omega_\lambda\right] F_{cv,\lambda}(\vec{k}) + \sum_{c'v'\vec{k}'} D_{c'v'\vec{k}'}^{cv\vec{k}} F_{c'v',\lambda}(\vec{k}') = 0, \quad (5)$$

with $\Delta_{cv}(\vec{k}) = E_c(\vec{k}) - E_v(\vec{k})$ being the single-particle energy differences between conduction and valence bands, and $\Omega_\lambda$ is the energy of the λ-th exciton state. $F_{cv,\lambda}(\vec{k})$ is the exciton envelope function and $D_{c'v'\vec{k}'}^{cv\vec{k}}$ is the direct Coulomb interaction taking into account the anisotropic potential due to the distinct dielectric environment in WZ ($\varepsilon_{xx} = \varepsilon_{yy} \neq \varepsilon_{zz}$ with values taken from Ref. 37). For the specific case in the presence of magnetic field, c and v indices run over the conduction and valence Landau levels [obtained with the Hamiltonian of equation (3)], the wavevector $k$ is two dimensional ($k_B$ and $k_A$ components) and the direct Coulomb term follows the same form as used in quantum well calculations, described in Refs. 32 and 33. Thus, the BSE is solved numerically for different values of magnetic field. Further information and a more technical description of the theoretical approach are discussed in section III B and IV of SM.

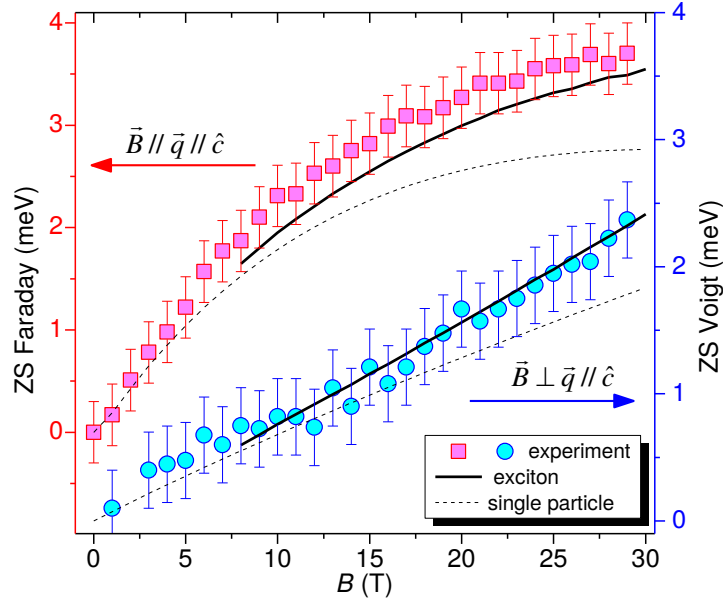

**Figure 3**. Exciton Zeeman splitting in the Faraday (squares, left axis) and Voigt (circles, right axis) configurations. Dashed lines represent the calculated values in the single-particle Landau levels at $k=0$. The solid lines are the calculated values with the inclusion of exciton effects. Note the double-ordinate and the explanation in note 38.

The calculated exciton ZS is shown in Figure 3 (solid lines) in comparison with the experimental data (colored symbols) and the single-particle Landau levels (dashed lines) at $\vec{k}=0$ (extracted from the data presented in Figure 2). For the Voigt configuration we also subtracted the electron-hole exchange energy that is not considered in the present theory. The single-particle curve taken at $\vec{k}=0$ does not take into account the effects of nonzero $k$-values, and therefore does not provide the full quantitative account of the experimental data but provides excellent qualitative agreement and a clear physical description. Instead, using the robust approach of the effective BSE for excitons, we are able to reproduce the experimental exciton ZS both in the Faraday and Voigt configurations very well. Contrasting with the single-particle curve at $\vec{k}=0$, the exciton calculations probe *via* Coulomb interaction nonzero $k$-values of the band structure that shows different ZS[38]. Ultimately, the Coulomb interaction averages (with particular weights obtained *via* the BSE) the contributions of different *k* points and consequently brings the values significantly closer to the experimental curve. We emphasize that remarkable agreement is attained without any free parameters. In addition, the same excellent agreement holds also for VLS InP NWs as reported in section V in SM. It is also worth mentioning that although the single-particle description does not provide the best quantitative agreement with the experimental data, it still offers a reliable qualitative behavior that could be used to explore these nonlinear features under different experimental conditions or in different material systems.

In conclusions, we presented a full description of the spin properties of WZ InP NWs under external magnetic field. The gyromagnetic factor of the exciton was determined by high-field magneto-photoluminescence using different orientations of the field with respect to the NW long symmetry axis. A markedly anisotropic $g_{exc}$ was found in accordance with the symmetry properties of the WZ lattice. In the Faraday configuration, the $g_{exc}$ value is about three times larger than in zincblende InP (see our values measured in a InP epilayer as described in section VI of SM) and therefore it becomes as high as in materials with quite large exciton *g*-factor, such as monolayer $WS_2$[39]. This increase in the *g*-factor is induced by the presence of the WZ phase found only in NWs, thus rendering WZ InP NWs a promising material for observing topological phases at reasonably low magnetic fields. Another difference between WZ and ZB InP that is important to point out is the absence of non linearity in the ZS of the ZB exciton. A linear behavior is indeed confirmed by our theoretical model, where due to HH-LH degeneracy in ZB all the couplings in the HH-LH block of the Hamiltonian are linear in magnetic field (more details can be found the SM section VI).

The theoretical model employed in this work reproduced the experimental Zeeman splitting data very well and this permitted us to highlight key aspects regarding the influence of the magnetic field on the electronic states of excitons in WZ NWs. First, admixing effects between Landau levels of the A and B valence bands occurred at fields as low as 5 T and induced a strong bowing of the holes' Zeeman splitting. Second, the inclusion of the electron-hole Coulomb interaction provided a realistic quantitative account of the experimental data due to the contribution of non-Γ states forming the exciton levels. These properties appear to be general features of WZ crystals and may play a relevant role in several WZ NW systems, where Majorana fermions are expected to become the building blocks of topological quantum computers.

AUTHOR INFORMATION

[*] corresponding author: antonio.polimeni@roma1.infn.it

[+]Present address: Department of Physics, University of Basel, Klingelbergstrasse 82, 4056 Basel, Switzerland

[⊥]Present address: Division of Physics and Applied Physics, School of Physical and Mathematical Sciences, Nanyang Technological University, Singapore 637371, Singapore

[×]Present address: Institute for Theoretical Physics and Astrophysics, University of Wuerzburg, 97074 Wuerzburg, Germany

*Acknowledgements*. We acknowledge the support of HFML-RU/FOM, member of the European Magnetic Field Laboratory (EMFL). Part of this work has been supported by EuroMagNET II under the EU contract number 228043. DT and MD acknowledge funding by Sapienza Università di Roma under the "Avvio alla Ricerca" grants. AP acknowledges financial support from "Awards 2014" and "Ateneo" funding schemes by Sapienza Università di Roma. A.G.D.A acknowledges the financial support of the Presidential Postdoctoral Fellowship program of the Nanyang Technological University. PEFJ and JF acknowledge the financial support of the Alexander von Humboldt Foundation, CAPES (Grant No. 99999.000420/2016-06) and SFB 1277 (B05). The Australian authors acknowledge the Australian Research Council for financial support and Australian National Fabrication Facility and Australian Microscopy and Microanalysis Research Facility for providing access to some of the equipment used in this work.

# Supplemental Material for the paper "Unusual spin properties of InP wurtzite nanowires revealed by Zeeman splitting spectroscopy"

D. Tedeschi,[1] M. De Luca,[1,2] P. E. Faria Junior,[3] A. Granados del Águila,[4] Q. Gao,[5] H. H. Tan,[5] B. Scharf,[3,6] P. C. M. Christianen,[4,7] C. Jagadish,[5] J. Fabian,[3] and A. Polimeni[1]

[1]*Dipartimento di Fisica, Sapienza Università di Roma, 00185 Roma, Italy*
[2]*Department of Physics, University of Basel, 4056 Basel, Switzerland*
[3]*Institute for Theoretical Physics, University of Regensburg, 93040 Regensburg, Germany*
[4]*High Field Magnet Laboratory (HFML - EMFL),*
*Radboud University, NL-6525 ED Nijmegen, The Netherlands*
[5]*Department of Electronic Materials Engineering, Research School of Physics and Engineering,*
*The Australian National University, Canberra, ACT 2601, Australia*
[6]*Institute for Theoretical Physics and Astrophysics,*
*University of Würzburg, 97074 Würzburg, Germany*
[7]*Division of Physics and Applied Physics, School of Physical and Mathematical Sciences,*
*Nanyang Technological University, Singapore 637371, Singapore*

## I. Polarization-resolved magneto-PL spectra

Figure 1 compares the PL spectra recorded under Faraday geometry for opposite spin-filtering. The $\Gamma_5^+$ components can be observed only for $\sigma^+$ circular light filtering, while for opposite circular polarization their contribution vanishes. The observation of the $\Gamma_5^-$ components for both polarizations is due to the thermally favored population of these lower-energy states.

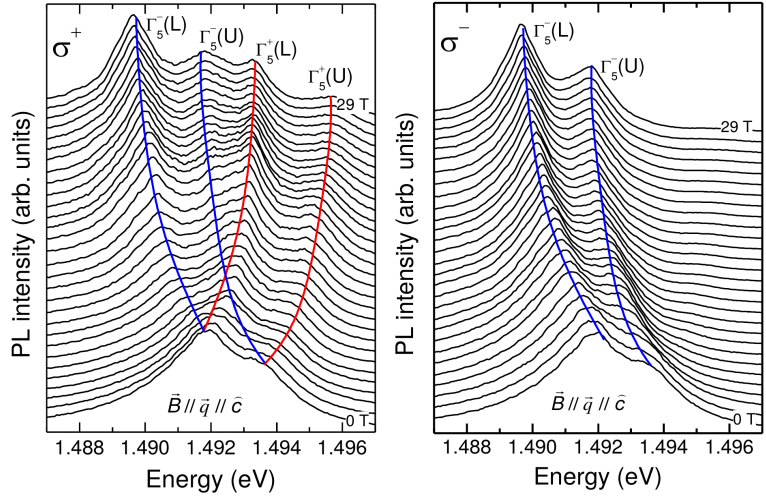

Figure 1: (Color online) Low-temperature (T=4.2 K) photoluminescence spectra of WZ InP NWs from 0 to 29 T (with increasing field from 0 to 29 T in step of 1 T from bottom to top). The relative orientation of magnetic field, emitted photon wavevector $\vec{q}$ and WZ $\hat{c}$ axis is $\vec{B} \parallel \vec{q} \parallel \vec{c}$. The spectra were filtered by $\sigma^+$ and $\sigma^-$ circular polarization in left and right panel, respectively. The different components are indicated and highlighted by colored lines. The energy axis of each $B \neq 0$ T spectrum has been corrected by the corresponding diamagnetic shift.

## II. Effective g-factors from theory

Within the theoretical framework of the k.p method, it is possible to calculate the values of effective g-factors for different energy bands[1–4]. The inclusion of external magnetic field in the bulk Hamiltonian with spin-orbit coupling (SOC) can be performed using the minimal coupling $\vec{p} \to \vec{p} + e\vec{A}(\vec{r})$ and adding the Zeeman term[4]. Since we are interested in the effect of uniform magnetic fields, we can choose[5] $\vec{A} = \frac{1}{2}\vec{B} \times \vec{r}$, which satisfies $\vec{\nabla} \cdot \vec{A} = 0$. In the resulting Hamiltonian, we can identify the following terms linear with the magnetic field $\vec{B}$ that leads to an effective Zeeman splitting (ZS):

$$H_{ZS} = \frac{\mu_B}{2} \vec{B} \cdot \left[ \underbrace{\frac{2m_0}{\hbar^2} \left( \vec{r} \times \vec{\Pi} \right)}_{\vec{\mathcal{L}}} + \underbrace{g_0 \vec{\sigma}}_{\vec{\mathcal{S}}} \right], \quad (1)$$

in which we have identified the two distinct contributions $\vec{\mathcal{L}}$ and $\vec{\mathcal{S}}$. Although the term $\vec{\mathcal{S}}$ is purely spin-dependent (composed by the Pauli matrices, $\vec{\sigma}$, and the bare electron g-factor, $g_0$), the term $\vec{\mathcal{L}}$ has both orbital and spin contributions due to the $\vec{\Pi}$ operator, given by:

$$\vec{\Pi} = \frac{\hbar}{m_0} \vec{p} + \frac{\hbar^2}{4m_0^2 c^2} \left[ \vec{\sigma} \times \vec{\nabla} V(\vec{r}) \right]. \quad (2)$$

Focusing on a spin-degenerate pair of bands at $\Gamma$-point of InP wurtzite (WZ) calculated by the k.p Hamiltonian of Ref. [6], i.e., conduction band $\{CB \Uparrow, CB \Downarrow\}$, heavy hole $\{HH \Uparrow, HH \Downarrow\}$, light hole $\{LH \Uparrow, LH \Downarrow\}$ or crystal-field split-off hole $\{CH \Uparrow, CH \Downarrow\}$, we can project the effective Zeeman Hamiltonian (1) in these eigenstates to extract their respective g-factors. Within this approach, each band is described by an effective Zeeman term of the form:

$$H_{ZS}^{\text{band}} = \frac{\mu_B}{2} g_\alpha^{\text{band}} B_\alpha \tau_\alpha, \ \alpha = x, y, z, \quad (3)$$

in which $\vec{\tau}$ are the Pauli matrices written in the $\Gamma$-point states and the effective g-factor $g_\alpha^{\text{band}}$ is obtained after evaluating the matrix elements of the $\vec{\mathcal{S}}$ and $\vec{\mathcal{L}}$ terms, given by

$$\vec{\mathcal{S}}_{nm} = g_0 \langle n | \vec{\sigma} | m \rangle \quad (4)$$

and

$$\hat{\alpha} \cdot \vec{\mathcal{L}}_{nm} = -i \frac{2m_0}{\hbar^2} \sum_{l \neq n,m} \frac{\langle n | \Pi_\beta | l \rangle \langle l | \Pi_\gamma | m \rangle - \langle n | \Pi_\gamma | l \rangle \langle l | \Pi_\beta | m \rangle}{E_n - E_l}, \quad (5)$$

with $\{\alpha, \beta, \gamma\} = \{x, y, z\}$, or cyclic permutations. To replace the $\vec{r}$ operator in the $\vec{\mathcal{L}}$ term, we used the relation

$$\langle a | \vec{r} | b \rangle = -i \frac{\langle a | \vec{\Pi} | b \rangle}{(E_a - E_b)}, \ E_a \neq E_b. \quad (6)$$

For the evaluation of expression (5), there are two points we should stress. First, it is helpful to notice that the matrix elements $\langle a | \vec{\Pi} | b \rangle$ are essentially the matrix elements of the Hamiltonians $H_{kp}^{(1)}$ and $H_{kSO}^{(1)}$, presented in Ref. [6]. Second, because of the interband SOC term $\Delta_4$ at $\Gamma$-point, the states for CB, LH and CH cannot be found analytically. Therefore, instead of carrying such numerical coefficients and deriving long expressions for the g-factors, we focus only on the numerical calculations to obtain the g-factors specifically for InP WZ. We note, however, that within the conventional 8×8 WZ (without the additional SOC terms), the expressions for CB g-factors follow the results of Hermann and Weisbuch[2].

In Table I we show the calculated values of the effective g-factors in the two magnetic field directions $B_x$ (Voigt) and $B_z$ (Faraday) for the energy bands at $\Gamma$-point CB, HH, LH and CH. As a general trend, $g_x$ (equal to $g_y$) and $g_z$ are different, highlighting the anisotropic feature of WZ structure, and except for CH band, we have $|g_z| > |g_x|$.

Table I: Calculated g-factors for the different energy bands of InP WZ. Due to the symmetry of the k.p Hamiltonian, we have $g_x = g_y \neq g_z$. In our notation, the $z$ ($x, y$) direction is parallel (perpendicular) to the c-axis of the WZ structure.

|       | CB   | HH    | LH    | CH   |
|-------|------|-------|-------|------|
| $g_x$ | 1.29 | 0.00  | -3.94 | 5.10 |
| $g_z$ | 1.61 | -3.05 | 5.12  | 0.47 |

## III. Landau levels

The Hamiltonian given by the equation (3) in the main text of the manuscript [with wavefunctions given in eq. (4)] that provides the Landau level (LL) physics of the system is essentially a set of coupled differential equations in real space which does not allow analytical solutions. We can solve it numerically by introducing finite differences approach[7] with constant material parameters. By choosing the vector potential with a single spatial dependence, the computational demand of the numerical calculations is greatly reduced. As a consequence of this choice in the vector potential form, we have two quantum numbers in $\vec{k}$ space: $k_B$, which is parallel to the magnetic field, and $k_A$, which is parallel to the vector potential. The spatial dependence of the vector potential is denoted by the coordinate $\rho$. In this notation of $k_B$, $k_A$ and $\rho$ (which are related to $x, y, z$ coordinates depending on the direction of magnetic field and choice of vector potential). The choices of vector potentials and directions of $k_B$, $k_A$ and $\rho$ for $B_x$ and $B_z$ are summarized in Table II. For the numerical calculations, we used a spatial length of $L = 200$ nm and $2N + 1$ discretization points with $N = 200$ (we define the space from $-L/2$ to $L/2$ with $N$ points to each side plus the point at zero). Therefore, we have roughly 1 point every 0.5 nm. We emphasize that within the envelope function approximation, the terms in the Hamiltonian are naturally symmetrized. Furthermore, it is important to notice that some second order terms in $k$ might need to be treated as first order derivatives due to the coupling with the vector potential, for instance $k_\alpha k_\beta \sim \frac{\partial}{\partial \alpha} A_\beta(\alpha)$.

Table II: Notation for the direction of applied magnetic field, the choice of vector potentials and the k vectors.

| $\vec{B}$ | $\vec{A}$ | $k_x$ | $k_y$ | $k_z$ | $k_B$ | $k_A$ | $\rho$ |
|---|---|---|---|---|---|---|---|
| $B\hat{x}$ | $By\hat{z}$ | $k_x$ | $-i\partial_y$ | $\frac{e}{\hbar}By + k_z$ | $k_x$ | $k_z$ | $y$ |
| $B\hat{z}$ | $Bx\hat{y}$ | $-i\partial_x$ | $\frac{e}{\hbar}Bx + k_y$ | $k_z$ | $k_z$ | $k_y$ | $x$ |

### A. Nonlinear features in Faraday configuration

In the Fig. 2 of the main text of the manuscript we showed that the nonlinear features of the ZS happen in the valence band due to the peculiar mixing of HH and LH LLs with different indices. Particularly for the Faraday configuration (B along along the NW axis) our experimental data (Fig. 1(c) of the main text) shows that the nonlinear features appear to only one type of light polarization. Our calculations support such findings by showing that indeed the nonlinearity is spin-selective, appearing only to the LL branch resulting from the couple of HH and LH with spin up states.

In order to provide additional insight on this finding, we can understand the couplings that give rise to the nonlinear features for the top LLs by looking at the WZ k.p Hamiltonian and at the composition of HH and LH states. In bulk InP, HH$\Uparrow$($\Downarrow$) is composed purely by the $|c_1\rangle$ ($|c_4\rangle$) basis state and LH$\Uparrow$($\Downarrow$) is composed mainly, $\sim 80\%$, by the $|c_2\rangle$ ($|c_5\rangle$) basis state. Furthermore, HH and LH states in WZ are split by an energy value of $\sim 35$ meV[6]. For this particular basis states that matter to our discussion, $|c_1\rangle$ and $|c_2\rangle$ are purely spin $\uparrow$, while $|c_4\rangle$ and $|c_5\rangle$ are purely spin $\downarrow$ states (for the complete basis set of the WZ Hamiltonian and their particular symmetry, please refer to Eq. 4 of Ref. [6]). The leading coupling between such states in the WZ k.p Hamiltonian is given by the second order k.p term (shown in the Appendix B of Ref. [6]) and it reads (leaving aside some constants)

$$\begin{aligned}\left\langle c_1 \left| H_{kp}^{(2)} \right| c_2 \right\rangle &\sim k_-^2 \\ \left\langle c_4 \left| H_{kp}^{(2)} \right| c_5 \right\rangle &\sim k_+^2,\end{aligned} \quad (7)$$

with $k_\pm = k_x \pm i k_y$.

In the framework of the envelope function approximation, the wavevectors $k_x$ and $k_y$ are the replaced by the operators given in Table II according to the choice of the magnetic field and choice of vector potential. For $\vec{B} = B\hat{z}$ and $\vec{A} = Bx\hat{y}$ it reads

$$\begin{aligned} k_x &= -i \frac{\partial}{\partial x} \\ k_y &= \frac{e}{\hbar} Bx. \end{aligned} \quad (8)$$

For the $k_-^2$ and $k_+^2$ operators, we can show that using the relations of Eq. (8) they can be written as the ladder operators ("creation", $a^\dagger$, and "annihilation", $a$) of the quantum mechanical harmonic oscillator so that:

$$\begin{aligned} k_-^2 &\sim (eBx + ip_x)^2 \sim a^2 \\ k_+^2 &\sim (eBx - ip_x)^2 \sim a^{\dagger 2} . \end{aligned} \quad (9)$$

Introducing these operators to the WZ k.p Hamiltonian in the envelope function approximation, we end up with the following selection rules between HH and LH LL-states

$$\begin{aligned} \langle f_{n,\text{HH}\Uparrow}(x) | a^2 | f_{m,\text{LH}\Uparrow}(x) \rangle &= \delta_{n,m-2} \\ \langle f_{n,\text{HH}\Downarrow}(x) | a^{\dagger 2} | f_{m,\text{LH}\Downarrow}(x) \rangle &= \delta_{n,m+2} , \end{aligned} \quad (10)$$

in which, for instance, $f_{n,HH\Uparrow}(x)$ is the envelope function with HH $\Uparrow$ character with $n$ nodes.

Focusing on the HH envelope function with $n = 0$, we verify from the selection rules presented in Eqs. (10) that the coupling between HH and LH states with spin up is achived for $m = 2$. On the other hand for spin down coupling between HH and LH states the requirement is $m = -2$, which cannot be satisfied since the number of nodes must be $\geq 0$. Therefore, only HH and LH states with spins up are coupled (HH wavefunction with 0 node and LH with 2 nodes) leading to the nonlinear dispersion observed in Fig. 2 of the main text. Such analysis with the quantum mechanical harmonic oscillator also holds if a different gauge is considered for the vector potential such as the symmetric gauge $\vec{A} = -By\hat{x}$ or even $\vec{A} = \frac{B}{2}(x\hat{y} - y\hat{x})$.

Such coupling between HH and LH states can be visualized in analogy to the optical orientation selection rules[8] in the bulk WZ structure, shown in Fig. 2(a). For the light polarization with a particular helicity ($S^-$, for instance) only spin up transitions from HH to CB are allowed. In order to activate the transitions for spin down, the helicity of the incoming light should be reversed. For the LL coupling depicted in Fig. 2(b), an analogous mechanism holds. Considering an applied magnetic field oriented along $\hat{z}$, only HH and LH states with spin up can be coupled and therefore give rise to the branch with nonlinear dispersion. Reversing the orientation of the magnetic field to $-\hat{z}$ would then allow the coupling between HH and LH states with spin down [we can see from Eq. (9) that by changing the sign of the magnetic field to $-B$, the role of the ladder operators is reversed so that now $k_-^2 \sim a^{\dagger 2}$ and $k_+^2 \sim a^2$].

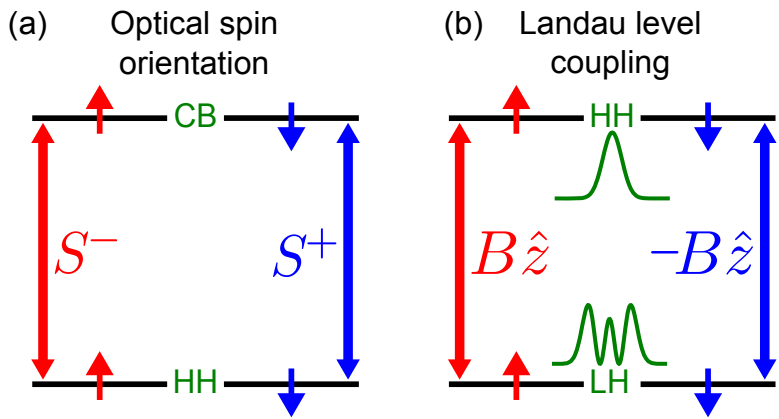

Figure 2: (Color online) Selection rules for (a) the optical orientation in bulk WZ materials for CB and HH bands and (b) the coupling of HH and LH states in the Faraday configuration, highlight the character of the HH and LH wavefunctions. The small arrows pointing up (down) represent the spin up (down).

### B. Band structure of Landau levels outside k=0

For the understanding of the origin of the nonlinear ZS features discussed in Fig. 2 of the main text and in the previous subsection here in the Supplemental Material, we can restrict our analysis to the energy levels and wavefunctions at $k_B = k_A = 0$. Although the main optical features follow the trends of $\vec{k} = 0$, we can expect additional contributions of nonzero $k$-points in order to obtain a more quantitative comparison to the experimental ZS. In Fig. 3 we show the energy dispersion of the LLs with respect to $k_B$ and $k_A$ for $B_x$ and $B_z$ at exemplary values of magnetic field of 10 T and 20 T. As we move away from the point $k_B = k_A = 0$, we observe the interplay of intrinsic SOC effects and magnetic field, which ultimately leads to a $k$-dependent ZS. To take into account the contributions of nonzero $k_B$ and $k_A$ quantum numbers in the ZS, we must deal with excitonic effects via the electron-hole Coulomb interaction, which couples conduction and valence band states via the Coulomb interaction at different $k$-points. The details for the inclusion of the excitonic effects is discussed in the next section of the Supplemental Material.

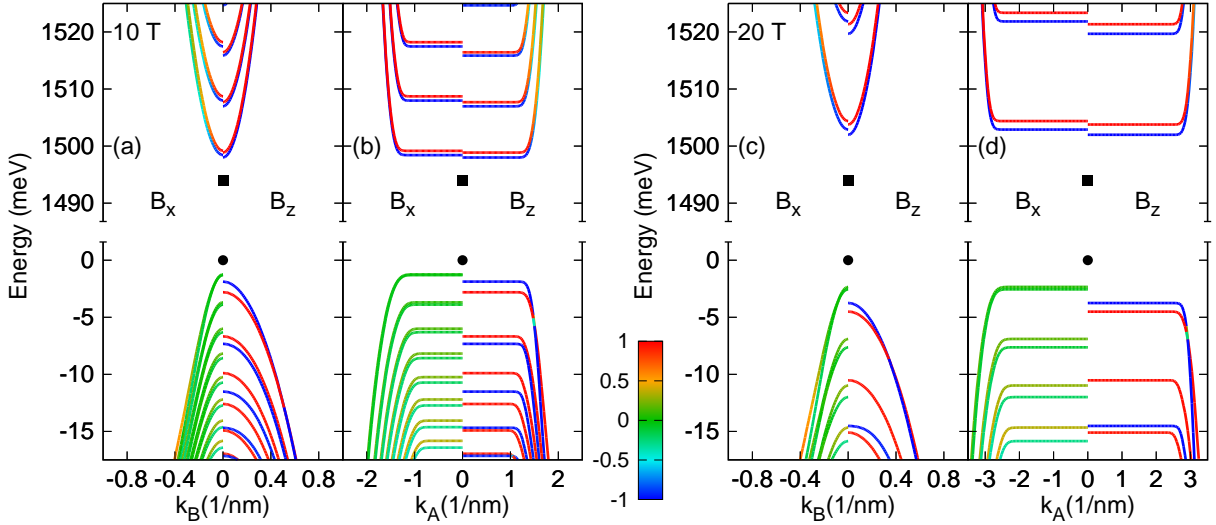

Figure 3: (Color online) Band structure at 10 T along (a) $k_B$ and (b) $k_A$ directions, and at 20 T along (c) $k_B$ and (d) $k_A$ directions. The points at $k_B = k_A = 0$ indicate the bulk energies at $\Gamma$-point for CB (square), HH (circle). In the x-axis, positive (negative) values indicate the applied magnetic field along $B_z$ ($B_x$). At $k_A$ direction, we notice the finite size effect, i. e., the deviation from a flat band dispersion. The color code indicates the spin orientation along the magnetic field.

### IV.  Excitonic effects

In this study, we are interested in electron-hole pairs optically generated by direct interband transitions at low temperature. Therefore, we focus on direct excitons with zero momentum at zero temperature. Formally, the mathematical treatment of excitons follows the effective BSE[9,10], given by the expression (4) in the main text of the manuscript which we repeat here for completeness:

$$\left[ E_c(\vec{k}) - E_v(\vec{k}) - \Omega_\lambda \right] F_{cv,\lambda}(\vec{k}) + \sum_{c'v'\vec{k}'} \mathcal{D}^{cv\vec{k}}_{c'v'\vec{k}'} F_{c'v',\lambda}(\vec{k}') = 0 \,, \tag{11}$$

in which $E_c(\vec{k})$ and $E_v(\vec{k})$ are the single-particle conduction and valence states (for instance, shown in Fig. 3 for B = 10 T and B = 20T), $\vec{k}$ is the wave vector given by $k_B$ and $k_A$ (according to Table II), $\Omega_\lambda$ is the energy of the $\lambda$-th exciton state with wave function given by

$$\Psi_\lambda(\vec{r}, \vec{r}') = \sum_{cv\vec{k}} F_{cv,\lambda}(\vec{k}) \psi_{c,\vec{k}}(\vec{r}) \psi^*_{v,\vec{k}}(\vec{r}') \,. \tag{12}$$

The direct Coulomb term, $\mathcal{D}^{cv\vec{k}}_{c'v'\vec{k}'}$, is generally written as

$$\mathcal{D}^{cv\vec{k}}_{c'v'\vec{k}'} = -\int d\vec{r} \int d\vec{r}' \, \psi^*_{c,\vec{k}}(\vec{r}) \psi_{v,\vec{k}}(\vec{r}') \mathrm{v}(\vec{r}, \vec{r}') \, \psi_{c',\vec{k}'}(\vec{r}) \psi^*_{v',\vec{k}'}(\vec{r}') \,, \tag{13}$$

with $\mathrm{v}(\vec{r}, \vec{r}')$ being the bare Coulomb potential for WZ materials, given by

$$\mathrm{v}(\vec{r}, \vec{r}') = \frac{e^2}{4\pi\varepsilon_0 \sqrt{\varepsilon_{xx}\varepsilon_{yy}\varepsilon_{zz}}} \left[ \frac{(x-x')^2}{\varepsilon_{xx}} + \frac{(y-y')^2}{\varepsilon_{yy}} + \frac{(z-z')^2}{\varepsilon_{zz}} \right]^{-\frac{1}{2}} \,, \tag{14}$$

with $\varepsilon_{xx} = \varepsilon_{yy} \neq \varepsilon_{zz}$ and values extracted from Ref. [11].

Before discussing the excitonic effects with magnetic field it is important to mention here that for the bulk case of InP WZ, we found an exciton biding energy of ∼7.6 meV considering only the HH band and estimated ∼9% decrease with inclusion of LH and CH bands. This theoretical result of ∼6.9 meV is in good agreement with the experimental estimation of 6.4 meV[12,13].

For the case with magnetic field, the direct Coulomb interaction can be expressed in a similar form as if we had a quantum well confinement[14–16], since both cases are treated via the EFA within the k.p method and have the quantum confinement along a single direction. We can then write $\mathcal{D}^{cv\vec{k}}_{c'v'\vec{k}'}$ using the single-particle wave functions [equation (4) of the main text], which reads

$$\mathcal{D}^{cv\vec{k}}_{c'v'\vec{k}'} = -\frac{e^2}{2\Lambda\varepsilon_0\varepsilon_\rho \gamma^{k_B k_A}_{k'_B k'_A}} \int d\rho \int d\rho' \, e^{-\gamma^{k_B k_A}_{k'_B k'_A}|\rho - \rho'|} \sum_{l=1}^{8} f^*_{c,k_B,k_A,l}(\rho) f_{c',k'_B,k'_A,l}(\rho) \sum_{m=1}^{8} f^*_{v',k'_B,k'_A,m}(\rho') f_{v,k_B,k_A,m}(\rho'), \tag{15}$$

with

$$\gamma^{k_B k_A}_{k'_B k'_A} = \sqrt{\frac{\varepsilon_B}{\varepsilon_\rho}(k_B - k'_B)^2 + \frac{\varepsilon_A}{\varepsilon_\rho}(k_A - k'_A)^2}. \tag{16}$$

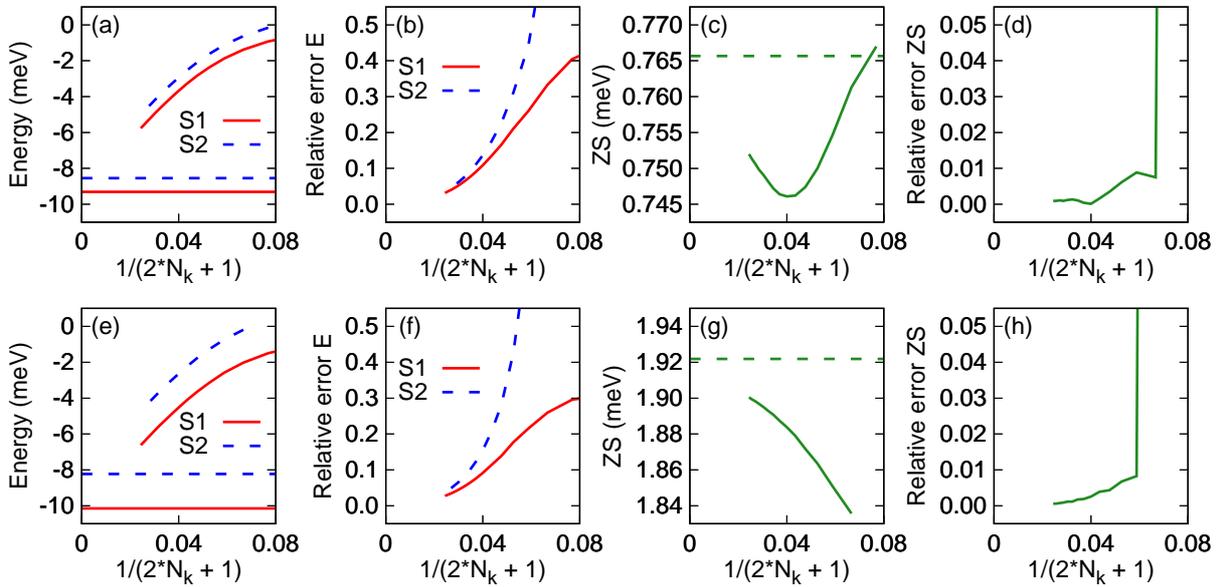

Figure 4: (Color online) Convergence of exciton energies with magnetic field as function of $1/(2*N_k+1)$ using the first Landau levels of conduction and valence band (totaling 4 states). (a) Exciton biding energy, (b) relative error of the energy, (c) Zeeman splitting and (d) relative error of the Zeeman splitting for $B_x$ configuration. (e-h) Same as (a-d) but for $B_z$ configuration. For this analysis we used $L = 100$ nm, $N = 50$, $B = 10$ T.

When dealing with excitons, convergence analysis with respect to the k-space and number of conduction and valence band states should be analyzed. Focusing on the two fundamental bright excitons, denoted here S1 and S2, that contribute with opposite light polarization, we show in Fig. 4 the convergence with respect to the number of k-points for the exciton biding energy, the relative error of the energy, the ZS and the relative error of the ZS for both magnetic field configurations ($B_x$ and $B_z$). For this analysis, we considered only the first LLs (top most in valence band and bottom most in conduction bands) and plotted the behavior with respect to $1/(2*N_k+1)$ with $N_k$ ranging from 1 to 20. We found that although the exciton biding energy decreases considerably as we increase $N_k$, which is necessary to achieve convergence and reduce the relative error, the exciton ZS is a robust property and converges much faster.

Now looking at the convergence with respect to the number of LLs, in Fig. 5 we show the convergence of the exciton biding energies and the ZS for $B_x$ and $B_z$ configurations at 10 T as function of the number of LLs in conduction and valence bands (each LL contains both spin configurations). We found that although the exciton biding energies change ∼1.7 meV, the ZS changes only 0.0638 meV for $B_z$ and 0.0408 meV for $B_x$. Furthermore, by increasing the strength of the magnetic field, the LLs become more energetically spaced, so that the contribution of additional LLs in the ZS should decrease even more. As a consequence, the excitonic effects of the nonlinear ZS can be investigated using the first LLs of conduction and valence bands. For this convergence analysis of the LLs, we considered $L = 100$ nm, $N = 50$ to reduce the time of the calculations since the two dimensional Coulomb integral of Eq. (15) is very time demanding. For the $k$-mesh, we used the same number of points for $k_B$ and $k_A$ directions: $N_k = 16$ with a total of $(2N_k + 1)^2$ points.

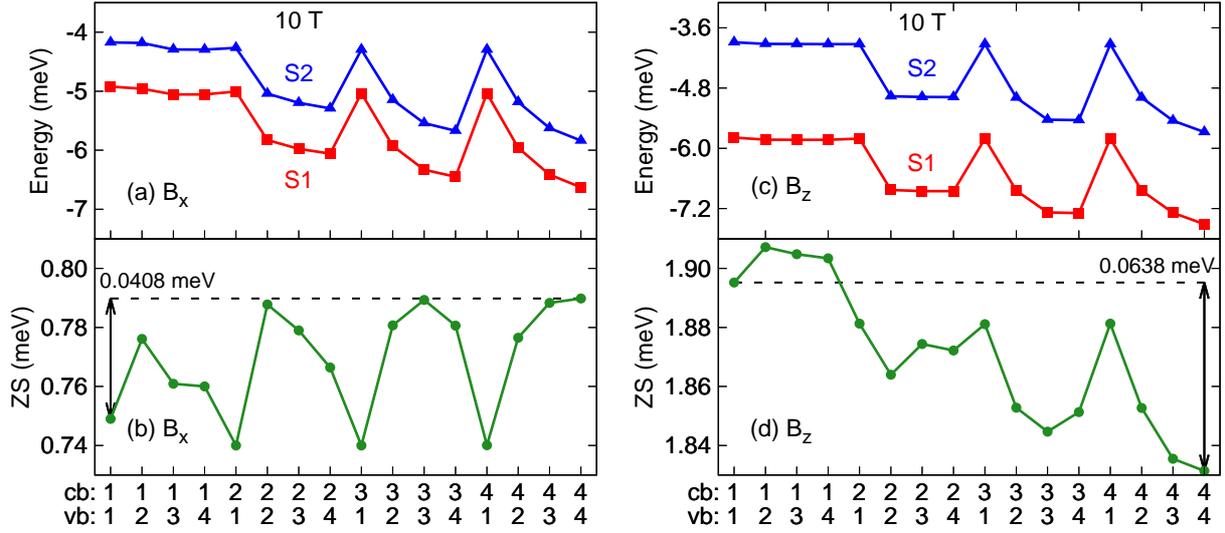

Figure 5: (Color online) Convergence of the exciton energies with magnetic field as a function of the number of Landau levels in conduction and valence bands. (a) Exciton biding energy and (b) Zeeman splitting for $B_x$ configuration. (c-d) Same as (a-b) but for for $B_z$ configuration. Each Landau level contains both spin contributions. For this analysis we used $L = 100$ nm, $N = 50$, $N_k = 16$ and $B = 10$ T.

For completeness, in Fig. 6 we compare the effect of $L = 100$ nm, $N = 50$ to $L = 200$ nm, $N = 200$ for the single particle energies at $\vec{k} = 0$ and for the exciton Zeeman splitting. Although $L = 200$ nm, $N = 200$ is closer to a realistic case, $L = 100$ nm, $N = 50$ still provides a reliable description and therefore can be used to estimate the convergence with respect to $N_k$ and LLs (as discussed in Figs. 4 and 5 here in the Supplemental Material) with significant reduced time in the calculations. The ZS was obtained by the linear extrapolation using $N_k = 19$ and $N_k = 20$, with the relative error between these two grid sizes being less than 3% for the exciton biding energies and nearly zero for the ZS.

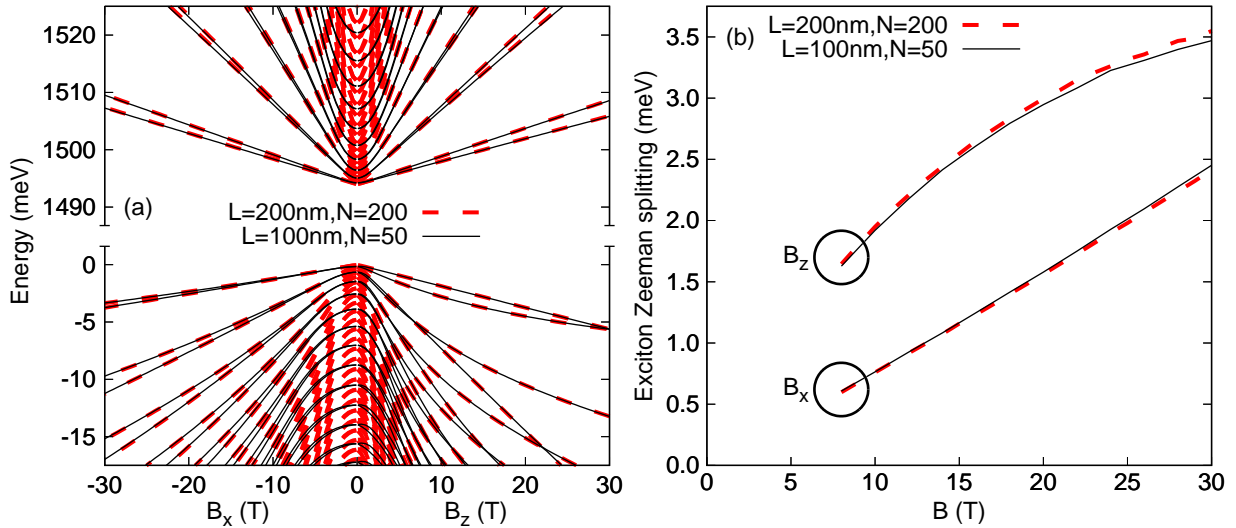

Figure 6: (Color online) Comparison of $L = 100$ nm, $N = 50$ and $L = 200$ nm, $N = 200$ for (a) the single-particle Landau levels and (b) the exciton Zeeman splitting using the 1st Landau levels of conduction and valence bands.

## V. Vapor-liquid-solid (VLS) vs selective area epitaxy (SAE) nanowire

In this section we show that the conclusions drawn in the manuscript for SAE InP NWs also hold for NWs formed via the VLS growth mode. The morphological characteristics of the two systems determined by scanning electron microscopy are displayed in Fig. 7(a). The SAE NWs have diameter and height equal to 650 nm and 5 $\mu$m, respectively. VLS Au-catalyzed NWs exhibits a tapered shape with approximate tip and base diameters of 30 nm and 200 nm, respectively. The NW length is about 2 mm and most of the luminescence originates from the NW base as detailed in Ref.[17].

Figure 7(b) compares the ZS in Faraday and Voigt configuration for SAE (smaller colored symbols; already reported in the main text) and VLS (larger gray symbols). The comparison with the theory for excitons is also shown (solid lines). The two sets of data agree very well between them.

Figure 8 shows the field dependence of the $\sigma^+$ and $\sigma^-$ components of the exciton ZS in the Faraday geometry for the VLS NWs. Similar to the SAE NWs (see Fig. 1 in the main text), the nonlinear behavior is due to the $\Gamma_5^+$ component.

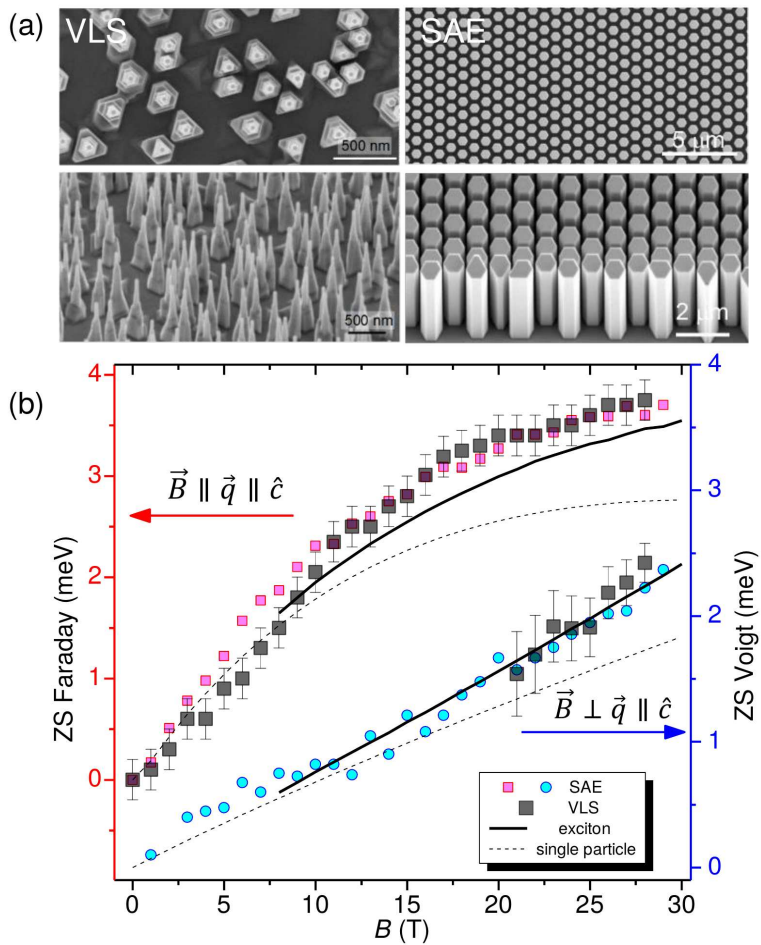

Figure 7: (Color online) (a) SEM images of the samples investigated. Left: top and side view of VLS NWs. Right: the same for SAE NWs. (b) ZS of SAE (smaller colored symbols), VLS (larger gray symbols) NWs and the theoretical calculations for single-particle (dashed lines) and excitons (solid lines). The ZS values are shown on the left and right axis for Faraday and Voigt geometry, respectively.

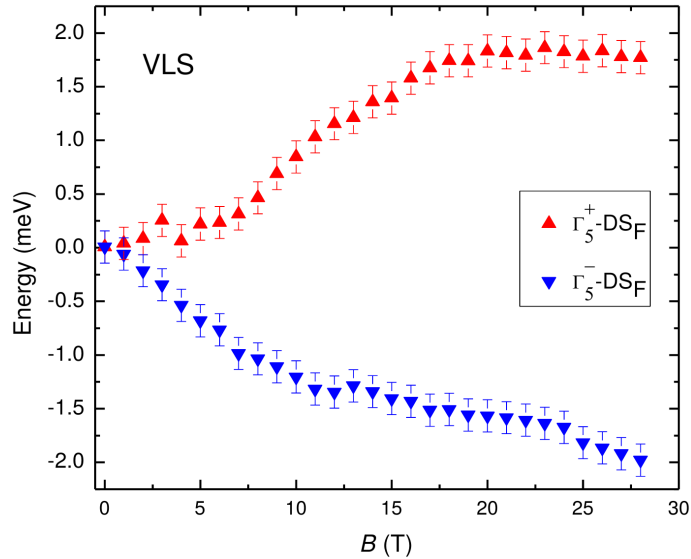

Figure 8: (Color online) Energy of the various Zeeman split free exciton components of VLS NWs in the Faraday geometry. The data have been obtained after subtraction of the diamagnetic shift, DS (see Ref. [18]). The same behavior shown in Fig. 1(c) of the main text for SAE NWs can be noted.

## VI. Zeeman splitting: WZ nanowires vs ZB bulk

In this section, we compare the extent and field dependence of the Zeeman splitting (ZS) in WZ NWs and in a bulk InP sample (ZB). This latter consists of a thick layer (3 $\mu$m) of InP grown epitaxially on an InP substrate similar to that on which the NWs were grown. Figure 9 shows the field dependence of the ZS in Faraday geometry ($\vec{B} \parallel \vec{q}$) for WZ NWs (same data shown in Fig. 3 of the main text) and ZB bulk InP. This latter exhibits a smaller ZS (resulting in an exciton gyromagnetic factor equal to 1.43) and the absence of nonlinear effects with magnetic field. In fact, by inspecting the HH-LH block in the conventional Luttinger-Kohn k.p Hamiltonian[19] using a similar approach of Sec. III A we can identify that all the couplings are linear in magnetic field and due to HH-LH degeneracy in ZB, a linear behavior is expected and indeed observed in Fig. 9.

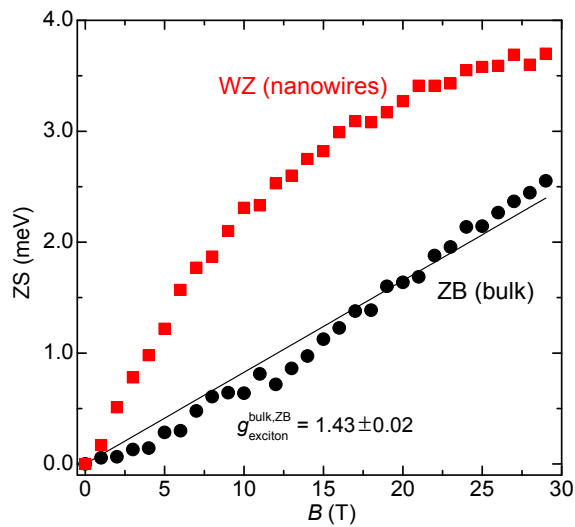

Figure 9: (Color online) Low-temperature (T=4.2 K) ZS of WZ InP NWs (red squares; same data as in Fig. 3 of the main text) and of bulk ZB InP (circles). The relative orientation of magnetic field and emitted photon wavevector $\vec{q}$ is $\vec{B} \parallel \vec{q}$. The solid line is a fitting to the ZB data resulting in an exciton gyromagnetic factor equal to 1.43.